\documentclass[prb,twocolumn,showpacs,amsmath,amssymb]{revtex4}

\usepackage{graphicx}
\usepackage{dcolumn}
\usepackage{bm}

\begin{document}

\title{ 
Kondo effect in asymmetric Josephson couplings 
through a quantum dot
}

\author{
Yoshihide Tanaka$^1$, Akira Oguri$^1$, and A. C. Hewson$^2$
}

\affiliation{
$^1$Department of Material Science, Osaka City University, 
Sumiyoshi-ku, Osaka 558-8585, Japan \\
$^2$
Department of Mathematics, Imperial College, 
180 Queen's Gate, London SW7 2BZ, UK
}

\date{\today}
 
\begin{abstract}
 Asymmetry in the Josephson couplings 
 between two superconductors through a quantum dot 
is studied based on a single impurity Anderson model 
using the numerical renormalization group (NRG).  
Specifically, we examine how the difference between
the couplings  $\Gamma_L$ and $\Gamma_R$
affects the ground state, which is known 
to show a quantum phase transition between 
a nonmagnetic singlet and a magnetic doublet 
depending on the various parameters; the Coulomb interaction $U$, 
  onsite potential $\epsilon_d$, level width $\Gamma_\mathrm{\nu}$ 
 caused by the hybridization, 
 and  superconducting gap $\Delta_\mathrm{\nu} 
 \equiv |\Delta_\mathrm{\nu}|e^{i \theta_{\nu}}$ 
 for the leads on the left 
 and right (${\nu} = L,\,R$).
Our results show that whether the local moment is fully 
screened or not depends substantially on the asymmetry in the couplings 
$\Gamma_L \neq \Gamma_R$. 
It tends to make the singlet ground state stable, 
while the amplitude $|\Delta_\mathrm{\nu}|$ 
and phase difference $\phi\equiv \theta_R-\theta_L$
of the superconducting gaps tend to suppress the screening.
We also discuss some general symmetry properties of the system  
and their relation to the current conservation. 
\end{abstract}

 \pacs{72.10.-d, 72.10.Bg, 73.40.-c}

\keywords{ Kondo effect, Josephson effect, 
 Quantum phase transition,  Anderson model, 
Numerical renormalization group
}


\maketitle

\section{Introduction}

The Kondo effect in superconductors has been studied over three decades.
\cite{SodaMatsuuraNagaoka,Matsuura,ShibaSoda,MullerHartmanZittartz}
It was shown in early years that 
the competition between the superconducting (SC) gap 
and Kondo energy scale $T_\mathrm{K}$ determines 
the low-temperature properties. 
The SC energy gap $\Delta_\mathrm{SC}$ disturbs 
the conduction-electron screening of the local moment.
The ground state becomes a magnetic doublet 
for $\Delta_\mathrm{SC} \gg T_\mathrm{K}$, 
while the ground state is still a nonmagnetic singlet 
for $\Delta_\mathrm{SC} \ll T_\mathrm{K}$.
These aspects of the Kondo physics in superconductors 
have been re-examined precisely by efficient numerical methods 
such as the quantum Monte Carlo (QMC) \cite{JarrelSiviaPatton}
and numerical renormalization group (NRG) approaches. 
\cite{SSSS,SSSS_2,YoshiokaOhashi,MatsumotoKoga}

In quantum dots embedded in a Josephson junction,
new interesting features have been predicted to be observed 
\cite{GlazmanMatveev,SpivakKivelson,Beenakker} 
and some experiments have been reported.  
\cite{Schoenenberger1,Schoenenberger2,Dam,Cleuziou}
Particularly, 
the screening of the local moment in the quantum dot  
is affected not only by the size of the two SC gaps 
but also by the Josephson phase $\phi$ between the 
two SC leads on the left and right.   
The phase difference $\phi$ also induces  
 the Josephson current flowing through the dot.
It can drive a quantum phase transition (QPT) 
between the singlet and doublet ground states, and 
at the critical point the direction of 
the current changes discontinuously. 

These properties of the SC-dot-SC systems have been 
studied with various theoretical approaches; such as  
the noncrossing approximation,\cite{Ishizaka,ClerkAmbegaokar} 
slave-boson mean-field theory,\cite{RozhkovArovas,Avishiai}
perturbation theory in the Coulomb interaction $U$,\cite{Vecino}
QMC,\cite{Kusakabe,siano_egger} and NRG.\cite{AO_YT_Hewson,Choi}
So far, however, 
most of the calculations have been carried out assuming a highly 
symmetric condition as 
$|\Delta_L| = |\Delta_R|$ and 
$\Gamma_L = \Gamma_R$. 
\cite{Ishizaka,ClerkAmbegaokar,RozhkovArovas,Avishiai,Vecino,siano_egger,Choi} 
Here, $\Delta_{\nu} \equiv |\Delta_{\nu}| 
\,e^{i \theta_{\nu}}$  
is the SC gap of the leads on the left 
and right (${\nu} = L,\,R$), 
and $\Gamma_\mathrm{\nu}$ is the bare level width due to the hybridization. 
Therefore, the feature of the QPT far from the symmetric condition  
is one of the theoretical issues to be clarified 
with some accurate approaches. 

In a previous work, we have studied the ground-state properties 
in the case of $|\Delta_L| \neq |\Delta_R|$.
In real quantum dots, it corresponds to the situation 
that the two superconductors connected to the dot 
are not identical.\cite{AO_YT_Hewson} 
Specifically, we have considered the limit  
where one of the two SC gaps, $\Delta_L$, 
is much larger than the other one $\Delta_R$.  
In the limit of $|\Delta_L|\to \infty$, 
the left lead has been shown to be separated from 
the rest of the system, which consists of the dot and right lead.  
It just leaves a static SC pair potential, which is given by 
 $\Delta_d = \Gamma_L \,e^{i \theta_L}$, 
in the dot. The pair potential appears because 
a SC proximity from the left lead is allowed even in the large gap limit, 
while the quasi-particle tunneling into the left lead is prohibited. 
This situation is described by a 
single-channel Anderson model with an extra SC pair potential 
$\Delta_d$ in the impurity site. The simplified model possesses 
essential ingredients of the Kondo and Josephson effects,
and the phase difference $\phi \equiv \theta_R-\theta_L$ 
can be defined between $\Delta_d$ and $\Delta_R$.
We have studied the features of QPT applying the NRG to this model, 
and have confirmed, for example, 
that the Andreev bound state emerging close to 
the Fermi level causes a re-entrant QPT near half-filling 
for $\phi \simeq \pi$ 
and $\Gamma_L \simeq \Gamma_R$.\cite{AO_YT_Hewson}

Another aspect that seems to be important is how the asymmetry 
in the couplings $\Gamma_L$ and $\Gamma_R$ 
affects QPT. That is the main subject of the present work. 
The SC proximity effect  
depends strongly on the environment surrounding the dot, 
and the competition between the Kondo and proximity effects determine 
the low-energy properties. In this paper we present the 
results of the QPT phase diagram obtained 
for the asymmetric couplings $\Gamma_L \neq \Gamma_R$.  
The NRG calculations have been carried out mainly 
using the two-channel model,
assuming the two gaps are equal 
in size $|\Delta_L| = |\Delta_R|$.
Our results show that the asymmetry in the couplings 
 tends to enhance the screening of the local moment. 

The model and 
some symmetry properties of the Anderson impurity  
in the Josephson junctions are described in Sec.\ \ref{sec:Model},
based on a rotational symmetry in the pseudo-spin space.
The NRG approach and the results of the phase diagram for QPT 
obtained in some typical parameter regions are
discussed in Sec.\ \ref{sec:results}. 
We also discuss in Sec.\ \ref{sec:discussion} 
the roles of the asymmetry in two solvable cases: 
noninteracting case $U=0$, 
and a superconducting version of the atomic limit 
for $|\Delta_L|\to \infty$ and $|\Delta_R|\to \infty$ 
which can also be considered as one of the fixed points of NRG.
A summary is given in Sec.\ \ref{sec:summary}. 
We also describe some general properties 
deduced from the current conservation in the appendix.

\section{Model for a SC-dot-SC junction}
\label{sec:Model}

We describe in this section  
the model and some of the symmetry properties. 
Specifically, at half-filling 
for the Josephson junctions with $\phi=0$ or $\pi$, 
the system has a global U(1) symmetry in a pseudo-spin space. 
It enables us to map the SC leads onto 
the normal conductors with a staggered potential, 
as shown by Satori {\em et al}.\cite{SSSS}

\subsection{General formulation}

\begin{figure}[t]
\leavevmode 
\begin{center}
\includegraphics[width=0.7\linewidth]{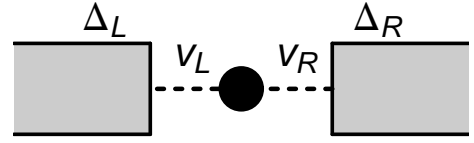}
\caption{
Schematic picture of an Anderson impurity connected to two SC leads 
with the gaps $\Delta_L$ and $\Delta_R$, 
where $v_L$ and $v_R$ are 
the tunneling matrix elements.
}
\label{fig:model}
\end{center}
\end{figure}

We start with  a single Anderson impurity connected 
to two SC leads on the left ($L$) and right ($R$), 
as illustrated in Fig.\ \ref{fig:model}.
The Hamiltonian is given by 
\begin{align}
\mathcal{H}  
\, =& \
 \mathcal{H}_{d}^0 
+ \mathcal{H}_{d}^U 
 +  \mathcal{H}_T 
+  \mathcal{H}_{c}^0 
+ 
 \mathcal{H}_{c}^{\mathrm{SC}} \;,
\label{eq:H}
\\
{\cal H}_{d}^0
=& 
\ \xi_d \, \bigl(  n_{d} -1 \bigr)
\;, \qquad 
{\cal H}_{d}^U = 
\frac{U}{2} \, \bigl(  n_{d} -1 \bigr)^2 , 
\label{eq:H_int}
\\
\mathcal{H}_T 
=&
\sum_{\nu=L,R} \sum_{\sigma} 
v_{\nu}^{\phantom 0}  
\left(\,
 c^{\dagger}_{\nu,0 \sigma} \, d^{\phantom{\dagger}}_{\sigma} 
\, + \,  
 d^{\dagger}_{\sigma} \,
 c^{\phantom{\dagger}}_{\nu, 0 \sigma}  
\, \right) ,
\label{eq:H_mix}
\\
\mathcal{H}_{c}^{0} 
=& 
\sum_{\nu=L,R} 
\sum_{n=0}^{\infty} \sum_{\sigma} 
t_{\nu,n}^{\phantom{0}} 
\left(
c^{\dagger}_{\nu, n+1\sigma}
c^{\phantom{\dagger}}_{\nu, n\sigma}
+
c^{\dagger}_{n, \nu \sigma} 
c^{\phantom{\dagger}}_{\nu,n+1\sigma}
\right) , 
\label{eq:H_c0}
\\
{\cal H}_{c}^{\mathrm{SC}} 
=&
\sum_{\nu=L,R} \sum_{n=0}^{\infty}  
\left(
\Delta_{\nu}^{\phantom{0}}\,
c^{\dagger}_{\nu,n\uparrow}\,
c^{\dagger}_{\nu,n\downarrow}  +   
\Delta_{\nu}^*\,
c^{\phantom{\dagger}}_{\nu,n\downarrow}
c^{\phantom{\dagger}}_{\nu,n\uparrow}
\right) .
\label{eq:sum_within_Debye}
\end{align}
Here, $\xi_d \equiv\epsilon_d + {U}/{2}$, 
with $\epsilon_d$ the impurity level,  
and $U$ is the onsite Coulomb interaction.  
The operator 
$d^{\dagger}_{\sigma}$ 
 creates an electron with spin $\sigma$ at 
the impurity site, 
and  $n_{d} = \sum_{\sigma}
d^{\dagger}_{\sigma} d^{\phantom{\dagger}}_{\sigma}$.  
The other operator 
 $c^{\phantom{\dagger}}_{\nu,0 \sigma}$ 
annihilates a conduction electron that 
 couples directly to the dot via the tunneling 
matrix element $v_{\nu}^{\phantom 0}$. 
The electrons in the leads 
can be described by 
the Hamiltonian of a tight-binding form 
without loss of generalities.\cite{Hewson}
The hopping matrix element $t_{\nu,n}^{\phantom{0}}$ determines 
a continuous one-particle spectrum $\epsilon_{k\nu}$ 
and eigenfunction $\varphi_{k\nu}^{\phantom{0}}(n)$,
by which  the hybridization strength can be written in the form 
$\Gamma_{\nu}^{\phantom{0}}(\varepsilon) = 
\pi v_{\nu}^2 \sum_k \left| 
\varphi_{k\nu}^{\phantom{0}}(0) \right|^2 
\delta(\varepsilon -\epsilon_{k \nu})$.

The symmetry properties of the system can be clearly seen,
in the Nambu representation using the spinors 
\begin{align}
\bm{\psi}_{d}^{\phantom{\dagger}} 
 = \!\! 
\begin{array}{l}
 \left[\,
 \begin{matrix}
  d_{\uparrow}^{\phantom{\dagger}} \cr
  d_{\downarrow}^{\dagger} \rule{0cm}{0.6cm}\cr
 \end{matrix}
 \,\right] 
\end{array} , \qquad 
\bm{\psi}_{\nu,n}^{\phantom{\dagger}} 
 = \! \!
\begin{array}{l}
 \left[
 \begin{matrix}
  c_{\nu,n\uparrow}^{\phantom{\dagger}} \cr
 (-1)^{n+1} c_{\nu,n\downarrow}^{\dagger} \rule{0cm}{0.6cm}\cr
 \end{matrix}
 \right] 
\end{array} .
\label{eq:spinor}
\end{align}
Then, 
 eqs.\ (\ref{eq:H_int})--(\ref{eq:sum_within_Debye})
can be rewritten in the form
\begin{align}
{\cal H}_{d}^0
=& 
\ \xi_d \  
\bm{\psi}_{d}^{\dagger}\, 
\mbox{\boldmath $\tau$}_3 \, 
\bm{\psi}_{d}^{\phantom{\dagger}} 
\;,
  \qquad 
{\cal H}_{d}^U = \frac{2U}{3}  \  
\vec{i}_d \cdot \vec{i}_d \;,
\label{eq:H_int_Nambu}
\\
\mathcal{H}_T 
=&
\sum_{\nu=L,R} 
v_{\nu}^{\phantom 0}  
\left(\,
\bm{\psi}_{\nu,0}^{\dagger}\,\bm{\psi}_{d}^{\phantom{\dagger}} 
+\bm{\psi}_{d}^{\dagger}\, \bm{\psi}_{\nu,0}^{\phantom{\dagger}} 
\, \right) ,
\label{eq:H_mix_Nambu}
\\
\mathcal{H}_{c}^{0} 
=& 
\sum_{\nu=L,R} 
\sum_{n=0}^{\infty} 
t_{\nu,n}^{\phantom{0}} 
\left(
\bm{\psi}_{\nu,n+1}^{\dagger}\,\bm{\psi}_{\nu,n}^{\phantom{\dagger}} 
+\bm{\psi}_{\nu,n}^{\dagger}\, \bm{\psi}_{\nu,n+1}^{\phantom{\dagger}} 
\right) , 
\label{eq:H_Nambu}
\\
{\cal H}_{c}^{\mathrm{SC}} 
=&
\sum_{\nu=L,R} \sum_{n=0}^{\infty}  
\, (-1)^{n+1} \  \bm{\psi}_{\nu,n}^{\dagger}\,
\mbox{\boldmath $\Delta$}_{\nu}\,
\bm{\psi}_{\nu,n}^{\phantom{\dagger}} 
\;.
\label{eq:H_sc_Nambu}
\end{align}
Here, $\mbox{\boldmath $\Delta$}_{\nu}$ 
and $\vec{i}_d$ 
are defined by
%
\begin{equation}
\mbox{\boldmath $\Delta$}_{\nu}
=  
\left[ \,
 \begin{matrix}
 0 &  \Delta_{\nu} \cr
 \Delta_{\nu}^* & 0
 \end{matrix}
 \, \right]  \;,
\qquad 
\vec{i}_d
= \frac{1}{2}\, 
\bm{\psi}_{d}^{\dagger}\, 
\vec{\mbox{\boldmath $\tau$}} \, 
\bm{\psi}_{d}^{\phantom{\dagger}} \;,
\end{equation}
and  
$\mbox{\boldmath $\tau$}_k$ for $k=1,2,3$ is 
the Pauli matrix. 
The Coulomb interaction ${\cal H}_{d}^U$ is isotropic 
in the pseudo-spin space, 
and the expression given in eq.\ (\ref{eq:H_int_Nambu}),
which is identical to that in eq.\ (\ref{eq:H_int}),
explicitly shows the rotational symmetry.
Furthermore, the current through the dot 
can be expressed in the form,
\begin{align}
  J_R  =& \,-i\,     
  \frac{e v_{R}^{\phantom{\dagger}}}{\hbar}
   \left(\, 
\bm{\psi}_{R,0}^{\dagger}\, 
\mbox{\boldmath $\tau$}_3 \, 
\bm{\psi}_{d}^{\phantom{\dagger}} 
-
\bm{\psi}_{d}^{\dagger}\, 
\mbox{\boldmath $\tau$}_3 \, 
\bm{\psi}_{R,0}^{\phantom{\dagger}} 
 \, \right) ,
 \label{eq:J_R_Nambu}
 \\ 
 J_L  =& \,-i\,     
  \frac{e v_{L}^{\phantom{\dagger}}}{\hbar} 
            \left(\, 
\bm{\psi}_{d}^{\dagger}\, 
\mbox{\boldmath $\tau$}_3 \, 
\bm{\psi}_{L,0}^{\phantom{\dagger}} 
-
\bm{\psi}_{L,0}^{\dagger}\, 
\mbox{\boldmath $\tau$}_3 \, 
\bm{\psi}_{d}^{\phantom{\dagger}} 
 \, \right) .
 \label{eq:J_L_Nambu}
\end{align}
The current is conserved such that  
 ${\partial n_d^{\phantom{0}}}/{\partial t} + J_R - J_L = 0$, 
as described also in the appendix \ref{sec:Green}.

The rotational symmetry in the pseudo-spin space 
are described by the operator 
\begin{align}
&
\vec{I} \, = \ 
\vec{i}_d
 \, +  \,  
\frac{1}{2}
\sum_{\nu=L,R}
\sum_{n=0}^{\infty}
\bm{\psi}_{\nu,n}^{\dagger}\,
\vec{\mbox{\boldmath $\tau$}} 
\,\bm{\psi}_{\nu,n}^{\phantom{\dagger}} 
 \;.
\end{align}
The $z$ component corresponds to 
the half of the total charge $Q=2I_z$, 
and $I_x$ ($I_y$) represents
the real (imaginary) part of the SC pair potentials. 
These three components satisfy the same commutation relations 
as that of the angular momentums,
 \begin{align}
 \left[ I_x,\, I_y \right] = i\,I_z 
 , \quad \ 
 \left[ I_y,\, I_z \right] = i\,I_x 
 , \quad \ 
 \left[ I_z,\, I_x \right] = i\,I_y 
 .
 \end{align}
In the pseudo-spin space,
the onsite potential $\xi_d$ can be interpreted 
as a local external field along the $z$ axis, 
and the SC gaps $\Delta_{\nu}^{\phantom{0}}$ can be 
regarded as a staggered field in the $x$-$y$ plane.  
Therefore, if $\xi_d=0$ and $\Delta_{\nu}=0$, 
which corresponds to normal leads in the particle-hole symmetric case, 
the system has a full rotational symmetry in the pseudo-spin space 
as well as the rotational symmetry in the real spin space, 
because $\left[I_{k},\,{\cal H}\right]=0$ for $k=x,\,y,\,z$.

Another typical case is that of the Josephson junction $\Delta_{\nu} \neq 0$ 
with the particle-hole symmetry $\xi_d=0$.
In this case the $z$ component of the external field vanishes.
Therefore, if we take the quantization axis to be in the $y$-direction 
by carrying out a spinor rotation of $\pi/2$ with respect to the $x$ axis,
the Hamiltonian $\mathcal{H}$ takes a real symmetric form. 
Then, with the new axis, the current operators are described by 
a Pauli matrix $\mbox{\boldmath $\tau$}_2$ 
instead of $\mbox{\boldmath $\tau$}_3$ 
in eqs.\ (\ref{eq:J_R_Nambu}) and (\ref{eq:J_L_Nambu}).

\begin{figure}[t]
\leavevmode 
\begin{center}
\includegraphics[width=0.7\linewidth]{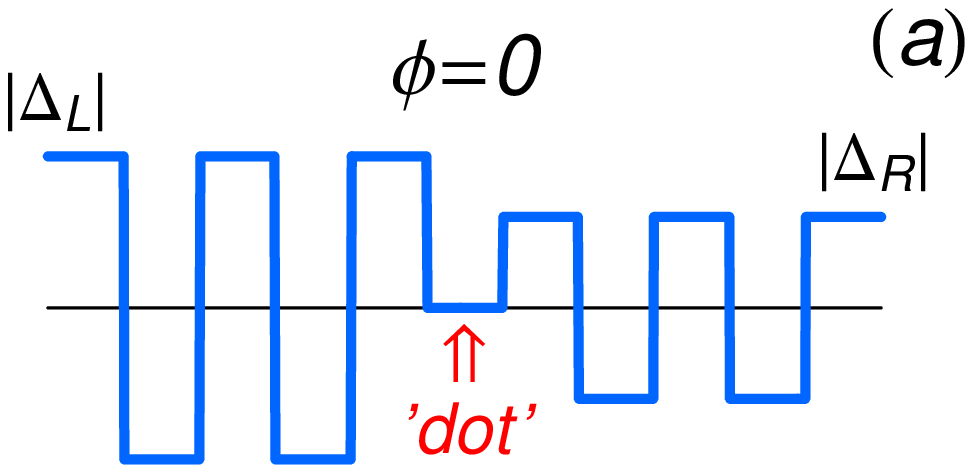}
\includegraphics[width=0.7\linewidth]{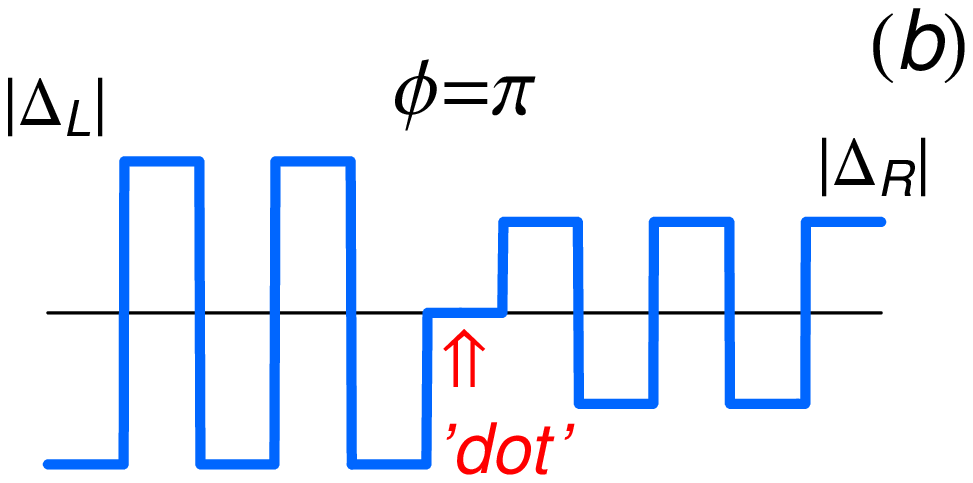}
\caption{
Potential profiles for the quasi-particles 
$\gamma_{\nu,n\sigma}$, given  
in eqs.\ (\ref{eq:gamma})--(\ref{eq:H_SC_gamma}),
for the Josephson junctions 
with (a) $\phi=0$, and (b) $\phi=\pi$, at half-filling $\xi_d=0$.
The Horizontal line corresponds to the Fermi level.
}
\label{fig:potentials}
\end{center}
\end{figure}

\subsection{Particle-Hole symmetric case with real gaps }

The two SC gaps  $\Delta_L$ and $\Delta_R$ 
are essentially real for 
the Josephson junctions with $\phi= 0$ or $\pi$.
In this case, if the system also has the 
particle-hole symmetry $\xi_d=0$, the model can be mapped onto 
a normal asymmetric Anderson model with 
a staggered potential.\cite{SSSS}
It is owing to a conservation associated 
with $\left[I_{x},\,{\cal H}\right]=0$, 
caused by a U(1) symmetry along the $x$ axis. 
It enables us to transform 
$I_x$ into the total number of the quasi-particles 
as shown in eq.\ (\ref{eq:N_gamma}). 
This feature can be seen explicitly   
if we take the quantization axis to be 
in the $x$ direction,\cite{SSSS} 
carrying out a spinor rotation of $\pi/2$ with respect 
to the $y$ axis by a unitary transform 
$e^{i\frac{\pi}{4}\bm\tau_2}$ as
\begin{align} 
\begin{array}{l}
 \left[
 \begin{matrix}
  \gamma_{-1\uparrow}^{\phantom{\dagger}} \cr
  \gamma_{-1\downarrow}^{\dagger} \rule{0cm}{0.6cm}\cr
 \end{matrix}
 \right] 
\end{array} 
\!\!\!
= \!
\left[ 
\begin{matrix} 
\frac{1}{\sqrt{2}}  &  \,  \frac{1}{\sqrt{2}} \cr
\frac{-1}{\sqrt{2}} \rule{0cm}{0.5cm}& \, \frac{1}{\sqrt{2}} \cr
\end{matrix} 
\right] 
\bm{\psi}_{d}^{\phantom{\dagger}} , \quad
 \begin{array}{l}
 \left[
 \begin{matrix}
  \gamma_{\nu,n\uparrow}^{\phantom{\dagger}} \cr
  \gamma_{\nu,n\downarrow}^{\dagger} \rule{0cm}{0.6cm}\cr
 \end{matrix}
 \right] 
\end{array} 
\!\!\!
=\!
\left[ 
\begin{matrix} 
\frac{1}{\sqrt{2}}  &  \,  \frac{1}{\sqrt{2}} \cr
\frac{-1}{\sqrt{2}} \rule{0cm}{0.5cm}& \, \frac{1}{\sqrt{2}} \cr
\end{matrix} 
\right] 
\bm{\psi}_{\nu,n}^{\phantom{\dagger}} .
\label{eq:gamma}
\end{align} 
Then, the Hamiltonian can be expressed in the form   
\begin{align}
&\mathcal{H}_{d}^U
=
\ \frac{U}{2} \, 
\left( 
\sum_{\sigma}
 \gamma_{-1\sigma}^{\dagger}\, 
\gamma_{-1\sigma}^{\phantom{\dagger}}
 -1 \right)^2  ,
\\
&\mathcal{H}_T 
 =   
\sum_{\sigma}
 v_{\nu}
 \left(
  \gamma^{\dagger}_{\nu,0 \sigma}\,
\gamma^{\phantom{\dagger}}_{-1\, \sigma}
  + 
 \gamma^{\dagger}_{-1\,\sigma}\, 
\gamma^{\phantom{\dagger}}_{\nu,0\sigma}
 \right) ,
\\
&\mathcal{H}_{c}^0 
 = \!  
\sum_{\nu=L,R} 
\sum_{n=0}^{\infty} \,
\sum_{\sigma}
 t_{\nu,n}
\! \left(
  \gamma^{\dagger}_{\nu,n+1\sigma} 
\gamma^{\phantom{\dagger}}_{\nu, n\sigma}
  + 
 \gamma^{\dagger}_{\nu, n \sigma}  
\gamma^{\phantom{\dagger}}_{\nu,n+1\sigma}
 \right) ,
\\
&\mathcal{H}_{c}^{\mathrm{SC}} 
=\!
\sum_{\nu=L,R} 
\sum_{n=0}^{\infty} 
 (-1)^{n+1} \Delta_{\nu} \! 
\left(\!
\sum_{\sigma}
   \gamma_{\nu,n\sigma}^{\dagger} 
\gamma_{\nu,n\sigma}^{\phantom{\dagger}}
 -1\! \right) 
.
\label{eq:H_SC_gamma}
\end{align}
Note that $\mathcal{H}_{d}^0 =0$, because of   
the particle-hole symmetry, $\xi_d=0$. 
Therefore, the energy spectrum is determined 
by the quasi-particles, the total number of which is conserved 
 \begin{align}
2 I_x = &
\ \, q_{-1}^{\phantom{0}} \, + 
\sum_{\nu=L,R} 
\sum_{n=0}^{\infty} 
 \left(
\sum_{\sigma}
   \gamma_{\nu,n\sigma}^{\dagger} 
\gamma_{\nu,n\sigma}^{\phantom{\dagger}} 
 -1 \right) \;, 
\label{eq:N_gamma}
\\
q_{-1}^{\phantom{0}} \equiv&\, 
\sum_{\sigma}   \gamma_{-1\sigma}^{\dagger} 
\gamma_{-1\sigma}^{\phantom{\dagger}}
 -1 
\ = \ 
d_{\downarrow}^{\phantom{\dagger}}
d_{\uparrow}^{\phantom{\dagger}}   
+
d_{\uparrow}^{\dagger}
d_{\downarrow}^{\dagger}  \;.
\end{align}
Particularly, the local {\em charge\/} 
$\,q_{-1}^{\phantom{0}}$ corresponds to 
the SC pair correlation penetrated into the impurity site. 
The profile of the potential that the 
quasi-particles feel is illustrated 
in Fig.\ \ref{fig:potentials}. 
For the Josephson junction with $\phi=0$  
the potential still shows a regular staggered period 
around the dot, while for $\phi=\pi$ the steps of the 
potential changes like that in a defect, or dislocation. 
The potential profile suggests 
that the SC proximity effect, which causes  
a finite anomalous average $\langle q_{-1}^{\phantom{0}} \rangle$, 
is larger for $\phi=0$ than that in the case of $\pi$. 
The penetration of the SC correlations into the 
impurity site depends on the tunneling strength 
$\Gamma_{\nu}^{\phantom{0}}$ and the size of the SC gaps
$\Delta_{\nu}^{\phantom{0}}$. 
Specifically, for $\phi=\pi$ and 
$\Gamma_R^{\phantom{0}}=\Gamma_L^{\phantom{0}}$, 
the impurity level evolves to 
the Andreev bound state emerging just 
on the Fermi level situated in the middle of the energy gap.
In this case, if also the size of the two gaps are 
equal  $\left|\Delta_R^{\phantom{0}}\right|
=\left|\Delta_L^{\phantom{0}}\right|$, 
the SC correlations cancel out $\langle q_{-1}^{\phantom{0}} \rangle=0$ 
in the impurity site. It also holds in 
the particle-hole asymmetric case $\xi_d \neq 0$ 
as shown in the appendix \ref{sec:Green} based on 
the current conservation.

\subsection{Particle-Hole asymmetric case for $\phi=\pi$}

There is a different symmetry for a $\pi$-junction  
$\phi =\pi$, in the case 
of $\Gamma_R^{\phantom{0}}
= \Gamma_L^{\phantom{0}}$ 
and  $\left|\Delta_R^{\phantom{0}}\right| 
=\left|\Delta_L^{\phantom{0}}\right|$. 
It holds without the particle-hole symmetry,
and reveals, for instance, 
in the Nambu form of the Green's 
function $\mbox{\boldmath $G$}$  given 
in eq.\ (\ref{eq:Green_pi}) in the appendix \ref{sec:Green}. 
The underlying symmetry links to 
the conservation of a channel charge $Q_\mathrm{ch}$
defined by
\begin{align}
&Q_\mathrm{ch} = \ n_d + 
\sum_{n=0}^{\infty} 
\sum_{\sigma}
 \left(
   a_{n\sigma}^{\dagger} a_{n\sigma}^{\phantom{\dagger}} 
 -  
   b_{n\sigma}^{\dagger} b_{n\sigma}^{\phantom{\dagger}} 
\right) \;, 
\label{eq:Q_ch}
\nonumber \\
&a_{n\sigma}^{\phantom{\dagger}} =  
\frac{
c_{R,n\sigma}^{\phantom{\dagger}} + c_{L,n\sigma}^{\phantom{\dagger}} 
}{\sqrt{2}}
 \;,\quad
b_{n\sigma}^{\phantom{\dagger}} =  
\frac{
c_{R,n\sigma}^{\phantom{\dagger}} - c_{L,n\sigma}^{\phantom{\dagger}} 
}{\sqrt{2}}
 \;.
\end{align}
The commutation relation,
 $\left[Q_\mathrm{ch},\,{\cal H}\right]=0$, 
can be confirmed explicitly if we express  
eqs.\ (\ref{eq:H_mix})--(\ref{eq:sum_within_Debye})
in terms of bonding and antibonding orbitals 
for $v_L=v_R$, and taking $t_{\nu,n}$ to be channel independent. 
The superconducting part, eq.\ (\ref{eq:sum_within_Debye}), 
 conserves the difference between 
the total number of the particles 
with the even symmetry and that of the odd ones, 
\begin{align}
\!\!\!\!
{\cal H}_{c}^{\mathrm{SC}} 
\! =\! 
\sum_{n=0}^{\infty}  
\Delta 
 \left(\!
   a_{n\downarrow}^{\phantom{\dagger}} b_{n\uparrow}^{\phantom{\dagger}} 
 +  
   b_{n\downarrow}^{\phantom{\dagger}} a_{n\uparrow}^{\phantom{\dagger}} 
+
b_{n\uparrow}^{\dagger} 
   a_{n\downarrow}^{\dagger} 
 +  
a_{n\uparrow}^{\dagger} 
   b_{n\downarrow}^{\dagger} 
\!
\right)  \! . 
\end{align}
Here, $\Delta$ is real and defined such that 
$\Delta_R^{\phantom{0}} =-\Delta_L^{\phantom{0}}=\Delta$. 
The Hamiltonian can be simplified carrying out    
the particle-hole transformation for $b_{n\sigma}^{\phantom{\dagger}}$,
\begin{align}
g_{n\uparrow}^{\phantom{\dagger}}
= \, (-1)^{n+1} \, b_{n\downarrow}^{\dagger}
\;, \qquad
g_{n\downarrow}^{\phantom{\dagger}}
= \,- (-1)^{n+1} \, b_{n\uparrow}^{\dagger} \;.
\end{align}
%
%
Then, the two leads consist a {\em normal\/} ladder
\begin{align}
&\mathcal{H}_{c}^0 
 = \,  
\sum_{n=0}^{\infty} \,
\sum_{\sigma}
 t_{n}
 \left(
  a^{\dagger}_{n+1\sigma} 
  a^{\phantom{\dagger}}_{n\sigma}
  + 
 g^{\dagger}_{n+1 \sigma}  
 g^{\phantom{\dagger}}_{n\sigma}
\, + \,  \mathrm{H.c.}  
\right) ,
\label{eq:Hc0_pi}
\\
&{\cal H}_{c}^{\mathrm{SC}} 
 =\, 
\sum_{n=0}^{\infty} \sum_{\sigma} \,
(-1)^{n+1} \, \Delta \left( 
   a_{n\sigma}^{\dagger} g_{n\sigma}^{\phantom{\dagger}} 
 +  
   g_{n\sigma}^{\dagger}  a_{n\sigma}^{\phantom{\dagger}} 
\right)  , 
\label{eq:Hsc_pi}
\end{align}
and one of the corners for $a_{0\sigma}^{\phantom{\dagger}}$ is 
connected to the dot as 
$\mathcal{H}_T 
 =   
\sum_{\sigma}
 \sqrt{2}v
 \bigl(
  a^{\dagger}_{0 \sigma}
d^{\phantom{\dagger}}_{\sigma}
 + 
 d^{\dagger}_{\sigma} 
a^{\phantom{\dagger}}_{0\sigma}
 \bigr)$.   
The dot part of the Hamiltonian,
 $\mathcal{H}_{d}^0+ \mathcal{H}_{d}^U$,  
keeps the original form given in eq.\ (\ref{eq:H_int}).
Thus, it is obvious that the SC correlation vanishes,  
$\langle d_{\uparrow}^{\dagger}d_{\downarrow}^{\dagger}\rangle=0$,
at the impurity site in the present situation even for 
the particle-hole asymmetric case $\xi_d \neq 0$.

The ladder described by 
eqs.\ (\ref{eq:Hc0_pi}) and (\ref{eq:Hsc_pi}) 
gives the density of states that shows 
a square root divergence at the edges 
of the gap $\epsilon=\pm \Delta$.
Therefore, the properties of the $\pi$-junction 
in the symmetric gaps and couplings must be 
essentially the same as those of the magnetic impurity 
in a band insulator, which has been studied using mainly 
a constant density of states with the gap around  
the Fermi level.\cite{Saso,Takegahara,ChenJaya}

\section{Ground-state properties}
\label{sec:results}

\subsection{NRG approach}

We study the ground-state properties using the NRG. 
In order to examine the effects caused by the asymmetry 
in the tunneling matrix elements, we consider the Josephson 
junction with $|\Delta_L| = |\Delta_R|$ ($\equiv \Delta$),
where the two gaps are equal in size.
To be specific, we assume that $\Gamma_{\nu}^{\phantom{0}}(\epsilon)$ 
is a constant inside the conduction band $-D< \epsilon <D$, 
and use the standard logarithmic discretization.\cite{KWW} 
Then, the hopping and tunneling matrix elements 
are expressed in the form 
\begin{align}
\,t_{\nu,n}^{\phantom{0}} 
&\,=\,   
D\,\frac{1+1/\Lambda}{2} \,
\frac{ \left(1-1/\Lambda^{n+1} \right) \ \Lambda^{-n/2}}
{\sqrt{1-1/\Lambda^{2n+1}}  \sqrt{1-1/\Lambda^{2n+3}} } 
\;,\\
 v_{\nu}^{\phantom{0}}
&\,=\, \sqrt{ \frac{2D\,\Gamma_{\nu} A_{\Lambda}}{\pi} }
\;,
\quad
A_{\Lambda}  \,=\,  \frac{1}{2}\, 
 \frac{1+1/\Lambda} {1-1/\Lambda }
\,\log \Lambda
\;.
\end{align}
The factor $A_{\Lambda}$ is required  
for reproducing the original model correctly 
in the continuum limit $\Lambda \to 1$.\cite{KWW,SakaiShimizuKasuya}
We have carried out the successive diagonalizations of NRG  
keeping typically the lowest $500$ energy states 
 for constructing the Hilbert space 
with extra orbitals from next step.
The low-energy properties are determined by 
the ratios $\Gamma_{\nu}/\Delta$, 
$U/\Delta$, $\epsilon_d/\Delta$ and the phase difference $\phi$. 
In the calculations we have taken the gap 
to be $\Delta/D=1.0 \times 10^{-5}$: 
this ratio itself does not affect 
the low-energy properties, provided $\Delta \ll D$.
We have used the eigenvalue of the pseudo spin $I_x$  
as a quantum number to classify the Hilbert space 
in the particle-hole symmetric case 
for $\phi=0$ and $\pi$. 
These two limits of $\phi$ provide 
the upper and lower bounds in 
the ground-state phase diagrams for intermediate $\phi$.

\begin{figure}[t]
\leavevmode 
\begin{center}
\includegraphics[width=0.95\linewidth, clip, 
trim = 1.0cm 16cm 1.0cm 0.0cm]{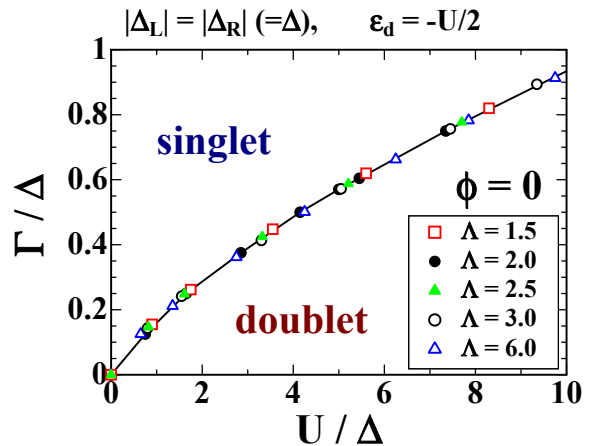}
%
\caption{
Phase diagram of the ground state for $\phi=0$ 
at half-filling $\epsilon_d=-U/2$
is plotted   
for several values of 
$\Lambda$: ($\square$) $1.5$, ($\bullet$) 
$2.0$, ($\blacktriangle$) $2.5$,
($\circ$) $3.0$, and ($\triangle$) $6.0$,
where 
 $\Gamma_L=\Gamma_R$ ($\equiv \Gamma$), and
$|\Delta_L| = |\Delta_R|$ ($\equiv \Delta$). 
}
\label{fig:fig90_SC_3}
\end{center}
\end{figure}

The model for the Josephson junction contains two conduction bands, 
and the accuracy of the NRG results becomes somewhat worse 
than that for a single-channel model.
For this reason, we have mainly used a rather large value 
$\Lambda=6.0$ for the discretization parameter 
to get an early convergence in the NRG steps. 
To check the validity of it,
we have examined the phase diagram of the ground state 
changing the value of $\Lambda$ ($=1.5,\, 2.0,\, 2.5,\, 3.0$ and $6.0$):  
 we have preserved the lowest $2000$ states for $\Lambda=1.5$.
The calculation has been carried out for $\phi=0$ 
where the model can be reduced to a single-channel model, 
and the results obtained at half-filling $\epsilon_d=-U/2$ for 
the symmetric couplings $\Gamma_L=\Gamma_R$ ($\equiv \Gamma$) 
are shown in Fig.\ \ref{fig:fig90_SC_3}.
We see that the results obtained at $\Lambda=6.0$ already 
agree well with those obtained for small $\Lambda$'s. 
It indicates that the level crossing of the ground and 
first excited states is not so sensitive to $\Lambda$, 
and it supports the validity of our results presented in the following.
The figure \ref{fig:fig90_SC_3} itself  
shows that the ground state is a nonmagnetic singlet 
for large $\Gamma$ or small $U$, and at the critical point 
it changes discontinuously to 
a magnetic doublet for small $\Gamma$ or large $U$.
\cite{AO_YT_Hewson,YoshiokaOhashi} 
\begin{figure}[t]
\leavevmode 
\begin{center}
\includegraphics[width=0.83\linewidth]{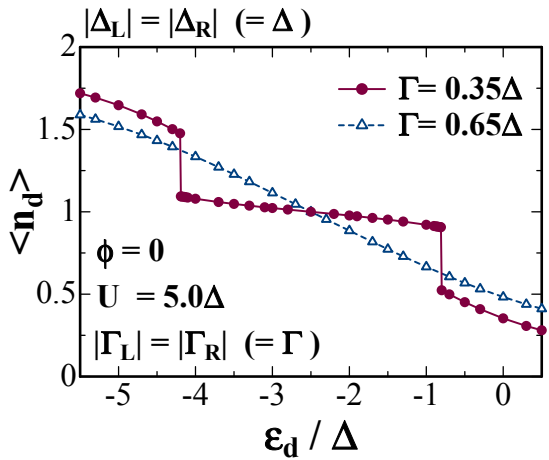}
\caption{
Number of electrons  in the dot $\langle n_d \rangle$ 
as a function of  $\epsilon_d$ for  $U=5.0\Delta$,
 $|\Gamma_L| = |\Gamma_R|$ ($\equiv \Gamma$), and 
$|\Delta_L| = |\Delta_R|$ ($\equiv \Delta$). 
The ground state is a singlet 
for ($\bullet$) $\Gamma = 0.35\Delta$, 
while for ($\triangle$) $\Gamma = 0.65\Delta$ 
it changes to a doublet for $-4.2 \lesssim \epsilon_d \lesssim -0.8$.
The discretization parameter is taken to be $\Lambda=2.0$.
}
\label{fig:fig83_SC}
\end{center}
\end{figure}

In order to show an example of   
the behavior of observables at the critical point, 
we have also calculated 
the average number of electrons $\langle n_d \rangle$ in the dot. 
The result is shown in Fig.\ \ref{fig:fig83_SC} as 
a function of $\epsilon_d$ 
for $U=5.0\Delta$,  $\phi= 0$, 
and  $\Gamma_L = \Gamma_R$. 
In this parameter set,  
the ground state for $\Gamma = 0.65\Delta$ is a singlet 
for all values of $\epsilon_d$, 
and thus $\langle n_d \rangle$ 
decreases monotonically 
with increasing $\epsilon_d$.
In contrast, for $\Gamma = 0.35\Delta$, 
the ground state is   
a doublet for $-4.2 \lesssim \epsilon_d \lesssim -0.8$,
and $\langle n_d \rangle$ jumps at the critical points. 
Other observable quantities, such as the Josephson current,  
will also show similar discontinuous changes, 
 because the QPT between the singlet 
and double ground states is the first order transition 
occurring as a result of a level crossing of 
the Andreev bound state emerging 
in the SC gap in the energy spectrum.

\begin{figure}[t]
\leavevmode 
\begin{center}
\includegraphics[width= 0.9\linewidth]{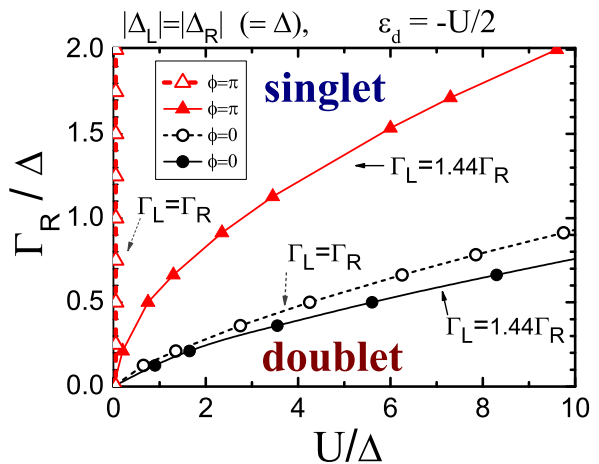}
\caption{Phase diagram of the ground state 
 at half-filling $\epsilon_d=-U/2$ 
for asymmetric couplings $\Gamma_L=1.44\Gamma_R$ 
with ($\bullet$) $\phi=0$ and ($\blacktriangle$) $\phi=\pi$.  
The dashed lines show the phase boundaries for the symmetric couplings
 $\Gamma_L=\Gamma_R$ with
 ($\circ$) $\phi=0$  and ($\triangle$) $\phi=\pi$.
The SC gaps are taken to be 
$|\Delta_L| = |\Delta_R|$  ($\equiv \Delta$).  
} 
\label{fig:fig92_SC}
\end{center}
\end{figure}

\subsection{Asymmetric couplings 
 $\,\Gamma_L \neq \Gamma_R\,$ at half-filling}

How does 
the asymmetry in the couplings $\Gamma_L$ and $\Gamma_R$ 
affect the ground-state properties?  
We first of all consider the particle-hole symmetric case $\epsilon_d=-U/2$. 
In this case, the dot is occupied by a single electron  
$\langle n_d\rangle =1$  on average, 
and magnetic correlations are enhanced maximally.
In Fig.\ \ref{fig:fig92_SC}, 
the phase diagram  of the ground state for
asymmetric couplings 
$\Gamma_L =1.44 \Gamma_R$ 
is plotted for the Josephson junctions with 
($\bullet$) $\phi=0$ and ($\blacktriangle$) $\phi=\pi$. 
For comparison, the results obtained for 
the symmetric couplings $\Gamma_L =\Gamma_R$ 
are shown for ($\circ$) $\phi=0$ 
 and ($\triangle$) $\phi=\pi$ with the dashed lines. 
The phase boundary for intermediate values of $\phi$,
 namely for $0<\phi<\pi$, 
will appear in between the line 
for $\phi=0$ and that for $\phi=\pi$.
In the figure, we see a general trend   
that the phase difference $\phi$ between the 
SC pair potentials  enlarges the region for  
a doublet ground state.\cite{AO_YT_Hewson} 
It means that the Josephson effects due to a finite $\phi$ disturbs 
the screening of the local moment, 
and the QPT will be driven by $\phi$ if 
the other parameters $U/\Delta$ and $\Gamma_{\nu}/\Delta$ bring the system 
into the parameter region enclosed by the two lines.
Note that for the data obtained  
for $\Gamma_L =1.44 \Gamma_R$, 
the sum of the hybridization energies $\Gamma_L+\Gamma_R$ 
is larger than that for the symmetric couplings. 
It also favors the singlet ground state.  

The results shown in Fig.\ \ref{fig:fig92_SC}  
suggest, however, that the superconducting 
proximity effects on the impurity site 
depends substantially on the asymmetry in the Josephson couplings,
particularly for the $\pi$-junction. 
The vertical phase boundary at $\phi=\pi$ 
for $\Gamma_L =\Gamma_R$ 
is caused by an Andreev bound state.
It emerges just on the Fermi level at half-filling,
as we see also in the potential 
profile shown in Fig.\ \ref{fig:potentials} (b).
Therefore, along the vertical line 
the ground state has 4-fold degeneracy.
An infinitesimal $U$ lifts the degeneracy  
when the two SC gaps are identical 
 $|\Delta_L|=|\Delta_R|$, 
and the ground state becomes a doublet in the whole region 
of $U>0$. We will discuss this point more in detail 
in Sec.\ \ref{sec:discussion}.   
The asymmetry in the couplings 
  $\Gamma_L \neq \Gamma_R$ breaks this situation.
Then, the Andreev bound state emerges away from the Fermi level,
and it makes the ground state to be a singlet for small $U$.

\begin{figure}[t]
\leavevmode 
\begin{center}
\includegraphics[width=0.92\linewidth]{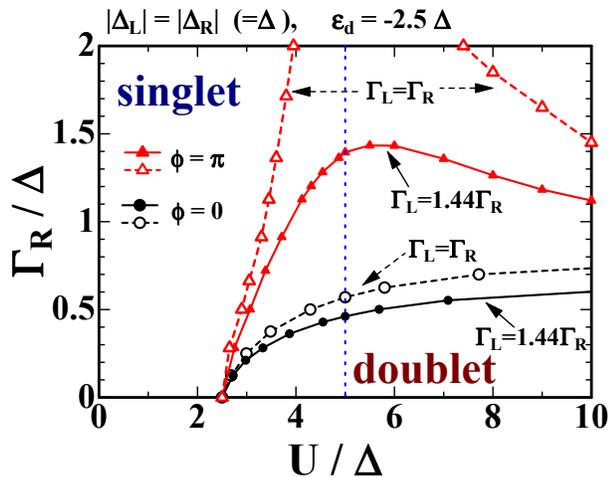}
\caption{
Phase diagram of the ground state away from half-filling 
at $\epsilon_d = -2.5\Delta$ is plotted for
 asymmetric couplings $\Gamma_L=1.44\Gamma_R$ 
for $(\bullet)$ $\phi=0$ and ($\blacktriangle$) $\phi=\pi$. 
Corresponding results for the 
symmetric couplings $\Gamma_L=\Gamma_R$ 
are also plotted for ($\circ$) $\phi=0$ and ($\triangle$) $\phi=\pi$. 
Particle-hole symmetry holds at $U=5.0\Delta$. 
}
\label{fig:fig80_SC}
\end{center}
\end{figure}

\subsection{Away from half-filling}

We also consider the ground state 
away from half-filling $\epsilon_d \neq U/2$.
The phase diagram obtained for a fixed value 
of the onsite energy $\epsilon_d = -2.5\Delta$ 
is shown in Fig.\,\ref{fig:fig80_SC}, 
where the particle-hole symmetry holds along the dashed line 
at $U=5.0\Delta$.
The calculations have been carried out 
for asymmetric couplings $\Gamma_L =1.44\Gamma_R$ 
for ($\bullet$) $\phi=0$ and ($\blacktriangle$) $\phi=\pi$,
and also for the symmetric couplings  
$\Gamma_L=\Gamma_R$ for  
 ($\circ$) $\phi=0$ and ($\triangle$) $\phi=\pi$.
As mentioned, 
the boundary for $0<\phi<\pi$ will appear between 
 the line for $\phi=0$ and that for $\phi=\pi$.
 In Fig.\,\ref{fig:fig80_SC}, all 
the boundaries start from the point $U=2.5\Delta$ 
in the horizontal axis. This is 
because in the atomic limit $\Gamma_{\nu}=0$  
the level crossing between 
the singlet and doublet states occurs 
at $2\epsilon_d+U =\epsilon_d$,
 when the energy of single-particle state and that of two-particle 
state coincide.
We also see in this figure that the 
phase boundary for $\phi =\pi$ is changed substantially  
by the asymmetry in the couplings, 
and the parameter region corresponding to 
the singlet ground state spreads in an upper-right region 
for $\Gamma_R \gtrsim 1.4 \Delta$ and $U \gtrsim 5.0 \Delta$. 
For the symmetric couplings, the phase boundary 
at $\phi=\pi$ approaches asymptotically to 
the vertical line at $U = 5.0\Delta$. 
This behavior is caused by the Andreev state 
approaching to the Fermi energy.

\begin{figure}[t]
\leavevmode 
\begin{center}
\includegraphics[width=0.9\linewidth]{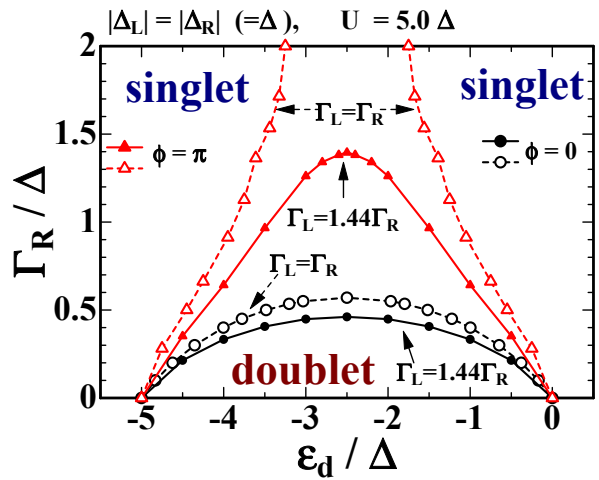}
\caption{
Phase diagram of the ground state away from half-filling 
at $U=5.0\Delta$ for    
 asymmetric couplings $\Gamma_L=1.44\Gamma_R$ 
for $(\bullet)$ $\phi=0$, ($\blacktriangle$) $\phi=\pi$, 
and the symmetric couplings $\Gamma_L=\Gamma_R$ 
for ($\circ$) $\phi=0$, ($\triangle$) $\phi=\pi$. 
Particle-hole symmetry holds at 
$\epsilon_d = -2.5\Delta$. 
} 
\label{fig:fig81_SC}
\end{center}
\end{figure}

In quantum dots  $\epsilon_d$ is a tunable parameter 
that can be controlled by the gate voltage.
The $\epsilon_d$ dependence of the 
phase diagram of the ground state for $\Gamma_L=\Gamma_R$ 
has been obtained first with the slave-boson mean-field 
approximation by Rozhkov and Arovas.\cite{RozhkovArovas}
We have examined the effects of the asymmetry.
The NRG results are shown in Fig.\,\ref{fig:fig81_SC} 
for (solid line) the asymmetric couplings
$\Gamma_L = 1.44\Gamma_R$, 
and for (dashed line) symmetric couplings 
$\Gamma_L=\Gamma_R$.
Here, the Coulomb interaction is fixed to be $U=5.0\Delta$.
The ground state is a singlet 
for $\epsilon_d+U<0$ and $\epsilon_d>0$, 
because the impurity site is almost fully filled  
or almost empty in these two regions.
We see again in Fig.\,\ref{fig:fig81_SC} 
that the asymmetry in the couplings
$\Gamma_L \neq \Gamma_R$ 
enlarges the parameter space of the singlet ground state. 
The phase boundary for $\phi=\pi$  
in the symmetric couplings $\Gamma_L=\Gamma_R$  
shows a diverging behavior at $\epsilon_d = -2.5\Delta$, 
which is owing to the Andreev bound state being situated  
just on the Fermi level at half-filling. 

\section{Discussions}
\label{sec:discussion}

The phase transition 
between the singlet and doublet ground states occurs 
at a level crossing of the lowest and the first 
many-body excited states, 
both of which are discrete and are separated 
from the continuous excitation spectrum 
at $\epsilon>\min(|\Delta_R|, |\Delta_L|)$ 
above the energy scale of the SC gap. 
Therefore, 
a few discrete states emerging inside the gap region 
determine the properties at low energy scales. 
In this section, to obtain some insights into 
how the asymmetry around the dot affects the ground states, 
 we consider two simple limits where the model can be solved.

 \begin{figure}[t]
 \leavevmode 
 \begin{center}
  \includegraphics[width=0.9\linewidth]{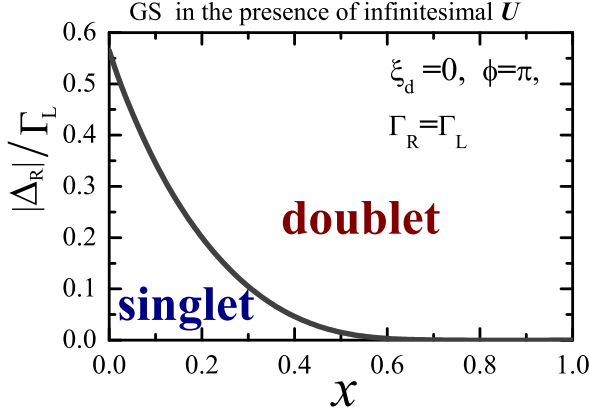}
 \caption{
Ground state for $\phi=\pi$, $\Gamma_R=\Gamma_L$, 
and $\xi_d=0$ in the presence of 
an infinitesimal Coulomb interaction $U$, 
which lifts the degeneracy caused by the zero-energy Andreev state. 
Here, $x = |\Delta_R|/|\Delta_L|$,  
represents the asymmetry in the SC gaps,
and $|\Delta_L|$ is changed in the region  
$|\Delta_L| > |\Delta_R|$ keeping $|\Delta_R|$ unchanged. 
The critical value of $\Gamma$ at $x=0$ is 
 $\Gamma_\mathrm{cr}\simeq 1.77|\Delta_R|$. 
 } 
 \label{fig:gamc}
 \end{center}
 \end{figure}

\subsection{Noninteracting limit $\,\mathcal{H}_d^U=0$}

We consider here how the asymmetry 
in the two SC gaps affects the Andreev state 
in the noninteracting case $\mathcal{H}_d^U=0$.
The density of states of each SC lead has 
the square root divergence at the edge 
 $\epsilon=\pm |\Delta_{\nu}^{\phantom{0}}|$,
and the formation of the bound state depends 
strongly on these divergences. The edges are separated from 
each other for $|\Delta_R^{\phantom{0}}| \neq |\Delta_L^{\phantom{0}}|$,
and they merge for the symmetric gaps 
to give a linear divergence for  $0<\phi\leq \pi$.

The energy of the Andreev state inside 
the gap is determined by an equation, 
$\det \left\{ \mbox{\boldmath $G$}^0(\epsilon) \right\}^{-1} =0$. 
The Green's function is defined in the appendix \ref{sec:Green}.  
The explicit form of the determinant 
 $F(\epsilon) \equiv 
\det \left\{ \mbox{\boldmath $G$}^0(\epsilon) \right\}^{-1}$ 
is given,  replacing the imaginary frequency $i\omega_n$ in 
eq.\ (\ref{eq:Gdd0_inv_explicit}) 
by the real one $\epsilon$ 
for $|\epsilon|< \min(|\Delta_R|, |\Delta_L|)$, as  
\begin{align}
F(\epsilon) =& 
\ \   \epsilon^2 - \xi_d^2   - \Gamma_L^2 - \Gamma_R^2 
\nonumber
\\
&    
+ \frac{2\Gamma_L \, \epsilon^2}{\sqrt{|\Delta_{L}|^2 - \epsilon^2 }} 
+ \frac{2\Gamma_R \, \epsilon^2}{\sqrt{|\Delta_{R}|^2 - \epsilon^2 }} 
 \nonumber
 \\
 &  + 
\frac{2\Gamma_L \Gamma_R  
\left( \epsilon^2-
|\Delta_{L}| |\Delta_{R}| \, \cos \phi \right)}{
 \sqrt{|\Delta_{L}|^2 - \epsilon^2} 
 \sqrt{|\Delta_{R}|^2 - \epsilon^2}  } 
 \,.
\label{eq:det_G0}
\end{align}
It takes a negative value, or vanishes, at $\epsilon=0$, 
\begin{align}
F(0)  
= - 
\xi_d^2 -  
\left(\Gamma_R^{\phantom{0}} +\Gamma_L^{\phantom{0}} \right)^2 
\cos^2 \frac{\phi}{2}
-\left(\Gamma_R^{\phantom{0}} -\Gamma_L^{\phantom{0}} \right)^2 
\sin^2 \frac{\phi}{2} \,.
\label{eq:det_G0_Fermi}
\end{align}
Therefore, the bound state emerges if $F(\epsilon)$ increases 
with $|\epsilon|$ to reach zero. 
For $\pi/2 \leq \phi \leq \pi$, the solution always exists 
because in this range of the phase 
$F(\epsilon)$ is an increasing function and 
diverges positively at $|\epsilon|=\min(|\Delta_R|, |\Delta_L|)$ 
with the square root dependence. 
Specifically, 
the bound state appears just on the Fermi level $\epsilon=0$,
when the right hand side of eq.\ (\ref{eq:det_G0_Fermi}) equals zero.
It happens for $\xi_d=0$, $\phi=\pi$, and  
$\Gamma_R^{\phantom{0}} =\Gamma_L^{\phantom{0}}$, 
provided $|\Delta_R^{\phantom{0}}|$ and 
$|\Delta_L^{\phantom{0}}|$ are nonzero. 
In this case the singlet 
and doublet ground states are degenerate 
because the energy does not depend on 
the way the bound state is occupied.
The asymmetry in the couplings 
 $\Gamma_R^{\phantom{0}}$ and $\Gamma_L^{\phantom{0}}$ affects 
crucially the ground-state property for the $\pi$-junction. 
The Coulomb interaction lifts the degeneracy, 
and it can be studied with a simple perturbation theory 
with respect to an infinitesimal $U$.\cite{AO_YT_Hewson}
The results are shown in Fig.\ \ref{fig:gamc},
assuming that $|\Delta_R|=\min(|\Delta_R|, |\Delta_L|)$.
The way how the degeneracy is lifted
depends quantitatively on the asymmetry in the SC 
gaps $x=|\Delta_R^{\phantom{0}}|/|\Delta_L^{\phantom{0}}|$. 
Specifically, the curve for the phase boundary reaches $0$ 
at $x=1$, which corresponds to the symmetric gaps 
$|\Delta_L^{\phantom{0}}|=|\Delta_R^{\phantom{0}}|$.
It means that the ground state is a doublet 
for any values of $\Gamma$ at $x=1$.
It explains the behavior of the vertical 
phase boundary for the $\pi$-junction in Fig.\ \ref{fig:fig92_SC}, 
and in this case the SC correlation in the impurity site vanishes 
$\langle d_{\uparrow}^{\dagger}  d_{\downarrow}^{\dagger} \rangle=0$ 
also for finite $U$ as shown in the appendix \ref{sec:Green}.

 \begin{figure}[t]
 \leavevmode 
 \begin{center}
 \includegraphics[width=0.9\linewidth]{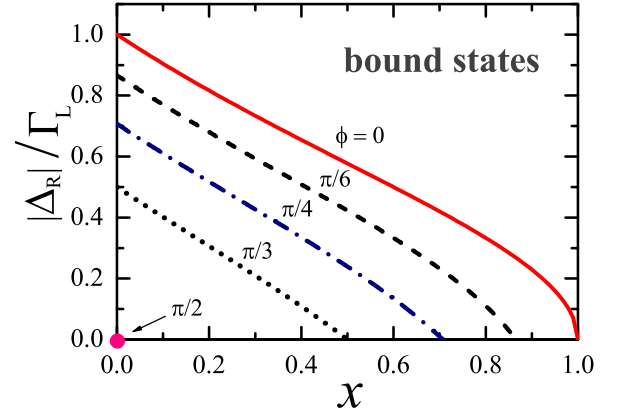}
 \caption{
 Bound states for $U=0$ {\em exist\/} in the region upside 
of each of the lines 
 for $\phi = 0,\, \pi/6,\,\pi/4$, and $\pi/3$.
  For $\pi/2 <\phi \leq \pi$, the bound states always exist.
 Here, $x = |\Delta_R|/|\Delta_L|$,  
 and $|\Delta_L|$ is changed in the region  
 $|\Delta_L| > |\Delta_R|$ keeping $|\Delta_R|$ unchanged.
The lines are determined by eq.\ (\ref{eq:bound_state}),
which does not depend on $\Gamma_R$ nor $\xi_d$. 
} 
  \label{fig:bound_state}
 \end{center}
 \end{figure}

 Another example seen in the case 
of asymmetric gaps $|\Delta_L^{\phantom{0}}| \neq |\Delta_R^{\phantom{0}}|$ 
is that the bound state disappears in a finite parameter region.
It is caused by the difference 
in the asymptotic behavior $F(\epsilon)$ 
at $|\epsilon|\to |\Delta_R^{\phantom{0}}|-0^+$, 
where 
we assume that $|\Delta_R^{\phantom{0}}| \leq |\Delta_L^{\phantom{0}}|$.
For instance, in the limit of $|\Delta_L^{\phantom{0}}| \to \infty$, 
the determinant $F(\epsilon)$ does not have a zero point 
for $|\Delta_R^{\phantom{0}}|<\Gamma_L^{\phantom{0}}\cos \phi$.
\cite{AO_YT_Hewson}
This is because the SC correlation of the order $\Gamma_L^{\phantom{0}}$, 
 which penetrates from left lead into the dot,  
pushed the bound state away towards the continuum outside of 
the gap region of the width $|\Delta_R^{\phantom{0}}|$. 
The bound state can remain inside 
the gap for $|\Delta_R^{\phantom{0}}|>\Gamma_L^{\phantom{0}}\cos \phi$.
The same situation also happens 
for finite $|\Delta_L^{\phantom{0}}|$.
The existence of the bound state 
can be deduced from the asymptotic behavior 
of the determinant at the edge of the gap region. 
The bound states appears 
if $F(\epsilon)\to +\infty$ 
 for $|\epsilon|\to |\Delta_R^{\phantom{0}}|-0^+$, 
and this condition can be expressed in the form, 
 for $0\leq x <1$ with $x = |\Delta_R|/|\Delta_L|$, as 
\begin{align}
\frac{|\Delta_R^{\phantom{0}}|}{\Gamma_L^{\phantom{0}}
}> \frac{\cos \phi -x}{\sqrt{1-x^2}} 
\;.
\label{eq:bound_state}
\end{align}
The solution does not exist 
when the both sides of eq.\ (\ref{eq:bound_state}) are equal, 
since $F(\epsilon)$ takes a negative finite value 
for $|\epsilon|\to |\Delta_R^{\phantom{0}}|-0^+$ 
at the critical condition. 
The parameter region where the bound sates 
do {\em not\/} exist shrinks as $\phi$ increases 
as shown in Fig.\  \ref{fig:bound_state}.
In contrast, for the symmetric gaps 
$|\Delta_R^{\phantom{0}}| = |\Delta_L^{\phantom{0}}|$, 
the determinant shows a positive linear divergence 
for $0<\phi\leq \pi$ at the edge of the gap, 
and the bound state always remains inside the gap region.

\subsection{Large gap limit $\,|\Delta_R|,|\Delta_L|  \to \infty$}

When both of the two gaps $|\Delta_R^{\phantom{0}}|$ 
and $|\Delta_L^{\phantom{0}}|$ are
larger than the other parameters, 
the model can be mapped onto a simple single site problem by taking 
a limit of $|\Delta_L^{\phantom{0}}| \to \infty$ 
and $|\Delta_R^{\phantom{0}}| \to \infty$.\cite{Affleck}
It can be regarded as a superconducting analogue of an atomic limit,
and can be solved analytically 
as summarized in the appendix \ref{sec:atomic_limit}.
The properties in the symmetric 
couplings $\Gamma_R^{\phantom{0}}=\Gamma_L^{\phantom{0}}$ 
have been studied by Vecino {\em et al}.\cite{Vecino}
In the following, we discuss mainly the role of the asymmetry 
in 
the couplings $\Gamma_R^{\phantom{0}}\neq \Gamma_L^{\phantom{0}}$. 

In the case of a normal transport through a quantum dot 
the equation of motion method,\cite{MWL} 
which is essentially the atomic limit 
in the Hubbard I approximation,\cite{HubbardI} 
has been applied to the properties at high energy scale. 
It has a limitation, however, at low temperatures 
because it does not describe properly the Fermi liquid properties 
due to the gapless low-energy excitations. 
Nevertheless, in the case of the dot connected 
to superconducting reservoirs, 
the low-energy excitations have the gap,  
and it gives a different meaning to the atomic limit.
The single site degrees of freedom can be 
identified as the Andreev bound state remaining inside the gap.
It does describe the lowest few many-body excited 
states inside the gap properly, and can also be considered 
as one the fixed points of NRG.
Although the virtual processes to the continuous excited states 
above the SC energy gap may also be important for 
the correlation functions, 
the main features of the contributions of the Andreev state 
to the low-energy properties, 
especially the bound-state energy itself, 
can be captured in the atomic limit 
of $|\Delta_L|$,$|\Delta_R| \to \infty$.

 \begin{figure}[t]
 \leavevmode 
 \begin{center}
 \includegraphics[width=0.8\linewidth]{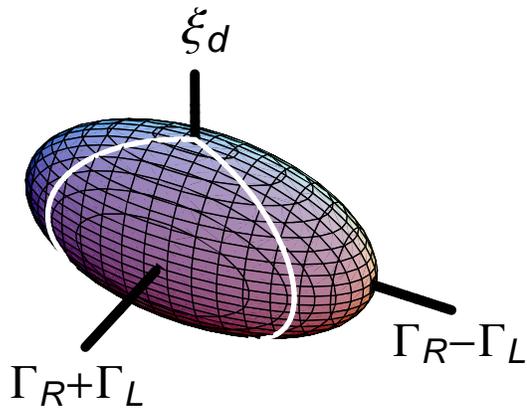}
 \caption{Equi-energy surface of Andreev state $E_A$ 
 in a parameter space consisting of $\Gamma_R$, $\Gamma_L$,
 and $\xi_d$ in the limit of $|\Delta_R|,|\Delta_L| \to \infty$. 
The phase $\phi$ changes the eccentricity of the 
ellipse in the horizontal plane, and  
the picture shows an example for a small $\phi$ ($\neq 0$).
The central region between the two white longitudinal lines, 
which represent $\Gamma_R=0$ and $\Gamma_L=0$, 
is the physical parameter space for $\Gamma_{\nu}>0$.
 The ground state is a doublet inside the surface 
which is determined such that $E_A=U/2$, 
and is a singlet for the outside.
 } 
 \label{fig:eg_surface}
 \end{center}
 \end{figure}

The effective Hamiltonian for the single site model 
is given in eq.\ (\ref{eq:H_atom}) in the appendix \ref{sec:atomic_limit}. 
In the large gap limit, the role of the SC leads can be 
replaced  by the SC pair potential at the impurity site
$\Delta_d = 
\Gamma_{L}^{\phantom{0}} e^{i \theta_{L}} 
\!+
\Gamma_{R}^{\phantom{0}} e^{i \theta_{R}}$, 
which is caused by the proximity effects.
The energy of the Andreev state 
$E_A =\sqrt{\xi_d^2 + |\Delta_d|^2}$ given in ($\ref{eq:E_A}$)
can be rewritten in the form, 
\begin{equation}
E_A 
=\sqrt{
\xi_d^2 + 
\left(\Gamma_R^{\phantom{0}} +\Gamma_L^{\phantom{0}} \right)^2 
\cos^2 \frac{\phi}{2}
+\left(\Gamma_R^{\phantom{0}} -\Gamma_L^{\phantom{0}} \right)^2 
\sin^2 \frac{\phi}{2}
} \;.
\label{eq:boundary_atomic1}
\end{equation}
Therefore, the equi-energy surface of $E_A$ can be described 
as an ellipsoid in a three-dimensional space 
consisting of 
$\xi_d$ and $\Gamma_R^{\phantom{0}} \pm \Gamma_L^{\phantom{0}}$
as shown in Fig.\ \ref{fig:eg_surface}.
The Josephson phase $\phi$ changes the aspect ratio, or eccentricity, 
of the ellipsoid.
Specifically, for $\phi \simeq \pi$, the gradient along the  
$\Gamma_R^{\phantom{0}} -\Gamma_L^{\phantom{0}}$ axis 
becomes large, and it implies that 
the asymmetry in the coupling plays an important role 
on the properties near the $\pi$-junction.

The ground state is a doublet for $E_A<U/2$, 
and it is a singlet for $E_A>U/2$.
Therefore, the Coulomb interacion determines the size of an ellipsoidal 
such that  
\begin{equation}
E_A \,= \,U/2 \;.
\label{eq:boundary_atomic2}
\end{equation}
The inside corresponds to a doublet 
ground state, and the outside is a singlet.  
The equations (\ref{eq:boundary_atomic1})
 and (\ref{eq:boundary_atomic2}) explain 
the features of the phase diagrams, shown in 
Figs.\  \ref{fig:fig92_SC}--\ref{fig:fig81_SC}, 
for small parameter values, at $\max(\Gamma_{R}^{\phantom{0}}, 
\Gamma_{L}^{\phantom{0}},U, |\xi_d|) \ll \Delta$. 
 Particularly, Fig.\ \ref{fig:fig81_SC} 
can be related to a cross section 
of the elliposoid for a fixed $U/2$.
For instance, for $\phi=\pi$, 
the phase diagram in the atomic limit is determined 
by $\xi_d^2 +(\Gamma_R^{\phantom{0}} -\Gamma_L^{\phantom{0}})^2 =U^2/4$, 
and thus the phase boundary becomes two open sheets 
at $\xi_d =\pm U/2$ for the symmetric couplings 
 $\Gamma_R^{\phantom{0}} =\Gamma_L^{\phantom{0}}$, 
while the asymmetry $\Gamma_R^{\phantom{0}} \neq \Gamma_L^{\phantom{0}}$ 
closes the sheets to form an elliptic cylinder. 
As another simple example, 
we consider the symmetric couplings
$\Gamma_R^{\phantom{0}} =\Gamma_L^{\phantom{0}}$ ($\equiv \Gamma$),
for which the boundary is  given  by 
$4\Gamma^2 \cos^2 \frac{\phi}{2} = U^2/4-\xi_d^2$.
This equation corresponds to the dashed lines shown in 
Figs.\ \ref{fig:fig92_SC} and  \ref{fig:fig80_SC} 
for small $\Gamma$.

In  the atomic limit  
the first and second many-body excitation energies 
$E_1$ and $E_2$, measured from 
the ground-state energy, are given by
\begin{align} 
& 
\mbox{singlet GS:}
&\!\!\!
\left\{ 
\begin{array}{l}
\displaystyle
E_1 = E_A - U/2  \\
\displaystyle 
E_2 = 2E_A 
\end{array}
\right.  \quad 
\mbox{for }  E_A > \frac{U}{2}\;,
\label{eq:from_singlet}
\\
& 
\mbox{doublet GS:}
& \!\!\! 
\left\{ 
\begin{array}{l}
\displaystyle
E_1 =  U/2 -E_A \\
\displaystyle 
E_2 = U/2 +E_A
\end{array}
\right.  \quad 
\mbox{for }  E_A < \frac{U}{2}\;.
\label{eq:from_doublet}
\end{align} 
Two discrete many-body excited states can appear 
in the energy gap even for finite SC gaps, 
for which the energies can be calculated with NRG. 
Then, eqs.\ (\ref{eq:from_singlet}) and (\ref{eq:from_doublet}) 
could be used to determine the renormalized values of $U$ and $E_A$,
which asymptotically coincide with the bare ones 
in the limit of  $|\Delta_L^{\phantom{0}}|, 
|\Delta_R^{\phantom{0}}| \to \infty$. 
The renormalization is caused  
by the contribution of the cotinuous states 
above the energy gap. 
The analysis of the fixed point 
along these lines will be discussed elsewhere.

\section{Summary}
\label{sec:summary}

We have studied effects of the asymmetry 
in the Josephson couplings $\Gamma_L \neq \Gamma_R$ 
on the ground state of a quantum dot embedded  between two superconductors. 
Specifically, we have obtained the phase diagram of the ground state 
in some parameter regions. The results show that the asymmetry 
in the couplings enhances the screening of 
the local moment in the dot, particularly for the $\pi$ junction. 
For $\phi \simeq \pi$, 
the Andreev bound state emerges near the Fermi level 
in the symmetric couplings  $\Gamma_L = \Gamma_R$,
and it favors the magnetic doublet ground state for $U>0$. 
The asymmetry in the couplings moves the Andreev bound state 
away from the Fermi level, and enhances the screening of the local moment. 
 The boundary between the singlet and doublet ground states
 for intermediate phase $0<\phi<\pi$ must be in between 
 the curves for $\phi=0$ and that for $\phi=\pi$.

We have also described some symmetry properties of 
the Anderson impurity in the Josephson junction 
in terms of the spinor rotation in the pseudo-spin 
and the current conservation. 
Specifically, the similarity between 
the $\pi$-junction and a magnetic impurity in 
an insulating host has been dicussed based on 
a symmetry property.
Furthermore, the role of the asymmetry 
in the couplings and gaps has been discussed 
in the noninteracting case, 
and also in the limit of  $|\Delta_L^{\phantom{0}}|,
 |\Delta_R^{\phantom{0}}| \to \infty$. 
Our results suggest that in real quantum dots 
the ground-state properties depends in various ways 
on the local conditions surrounding the dot.


\section*{Acknowledgements}
We would like to thank Yoichi Tanaka, Y.\ Nisikawa, 
and J.\ Bauer for valuable discussions.
 One of us (AO) is supported by the Grant-in-Aid 
for Scientific Research from JSPS. 
ACH wishes to thank the EPSRC(Grant GR/S18571/01) 
for financial support.
Numerical computation was partly carried out 
at the computation center of Yukawa Institute.

\appendix

\section{Current conservation}
\label{sec:Green}

The imaginary-time Green's function for the Anderson impurity 
in the SC hosts can be described in the Nambu representation as
\begin{align}
\!\!\!
\mbox{\boldmath $G$}(i \omega_n) 
 = - 
\!\!\int_0^{\beta} \!\!\!\! d\tau \, e^{i\omega_n \tau}
\! \left[ 
 \begin{matrix}
\langle T_{\tau}  
  d_{\uparrow}^{\phantom{\dagger}}(\tau) \,d_{\uparrow}^{\dagger}
\rangle 
& \!
\langle T_{\tau}  
  d_{\uparrow}^{\phantom{\dagger}}(\tau) \,d_{\downarrow}^{\phantom{\dagger}}
\rangle 
\cr
\langle T_{\tau}  
  d_{\downarrow}^{\dagger}(\tau) \, d_{\uparrow}^{\dagger}
\rangle 
& \!
\langle T_{\tau}  
 d_{\downarrow}^{\dagger}(\tau) \, d_{\downarrow}^{\phantom{\dagger}}
\rangle 
 \end{matrix}
 \right]  \!\! ,
\label{eq:Gdd}
\end{align}
where $\omega_n = (2n+1)\pi/\beta$ for $n=0,\pm1 ,\pm 2, \ldots$, 
$\, \beta=1/T$, and $T$ the temperature.
The Dyson equation is given by 
\begin{align}
\left\{ \mbox{\boldmath $G$}(i \omega_n) \right\}^{-1} 
= \, 
\left\{ \mbox{\boldmath $G$}^0(i \omega_n) \right\}^{-1} 
 -
\mbox{\boldmath $\Sigma$}(i \omega_n)
\;,
\label{eq:Gdd_inv}
\end{align}
where $\mbox{\boldmath $\Sigma$}(i \omega_n)$ is the self-energy 
caused by $\mathcal{H}_d^{U}$, and 
\begin{align}
&\!\!\!\!\!
\left\{ \mbox{\boldmath $G$}^0(i \omega_n) \right\}^{-1} 
\!= \,  i \omega_n \mbox{\boldmath $1$} 
-  \xi_d\,\mbox{\boldmath $\tau$}_3 
 + \!\sum_{\nu=L,R} \! \Gamma_{\nu}^{\phantom{0}}  
 \frac{ i \omega_n  \mbox{\boldmath $1$} - \mbox{\boldmath $\Delta$}_{\nu}}
 {\sqrt{\omega_n^2 + |\Delta_{\nu}|^2 } } .
\label{eq:Gdd0_inv_explicit}
\end{align}

The currents through the dot is defined 
based on the Heisenberg equation,  
${\partial n_d}/{\partial t} = \bigl(-i/ \hbar \bigr)
\bigl[ n_d,\, \mathcal{H} \bigr]$, as
\begin{align}
& 
e\,
\frac{\partial n_d^{\phantom{0}}}{\partial t} + J_R - J_L = 0 
\;,\\
 &  J_R  = \,-i\,     
  \frac{e v_{R}^{\phantom{\dagger}}}{\hbar}
 \sum_{\sigma}
   \left(\, 
            c^{\dagger}_{R,0 \sigma} 
            d^{\phantom{\dagger}}_{\sigma} 
         -  d^{\dagger}_{\sigma} 
            c^{\phantom{\dagger}}_{R,0  \sigma} 
 \, \right) ,
 \label{eq:J_R}
 \\ 
 &  J_L  = \,-i\,     
  \frac{e v_{L}^{\phantom{\dagger}}}{\hbar} 
 \sum_{\sigma}
            \left(\, 
             d^{\dagger}_{\sigma} 
             c^{\phantom{\dagger}}_{L,0 \sigma}  
           - c^{\dagger}_{L,0 \sigma} 
             d^{\phantom{\dagger}}_{\sigma}
 \, \right) .
 \label{eq:J_L}
\end{align}
Here, $J_L$ is the current flowing into the dot 
from the left,
$J_R$ is the current from the dot to the right.
The thermal average of these currents 
can be expressed in terms of the Green's function 
\begin{align}
\langle J_R \rangle 
 &=   \frac{2 e }{\hbar\beta} 
 \sum_{\omega_n}
\frac{ i\,\Gamma_R^{\phantom{0}}}
{\sqrt{\omega_n^2 + |\Delta_{R}|^2 }} \ 
\mbox{Tr}\Bigl[\,
\mbox{\boldmath $\tau$}_{3} \mbox{\boldmath $\Delta$}_{R} 
\mbox{\boldmath $G$}(i\omega_n)
\,\Bigr] , 
\label{eq:JR_Joseph}
\\
 \langle J_L \rangle  
 &=  \frac{2 e }{\hbar\beta} 
 \sum_{\omega_n}
\frac{-i\,\Gamma_L^{\phantom{0}}}
{\sqrt{\omega_n^2 + |\Delta_{L}|^2 }} \ 
\mbox{Tr}\Bigl[\,
\mbox{\boldmath $\tau$}_{3} \mbox{\boldmath $\Delta$}_{L} 
\mbox{\boldmath $G$}(i\omega_n)
\,\Bigr] .
\label{eq:JL_Joseph}
\end{align}
Here, the trace is taken over 
for the $2 \times 2$ matrices for the pseudo spin space. 
The current conservation 
 $\langle J_L \rangle = \langle J_R \rangle$ 
indicates that there is one relation between the off diagonal 
elements of $\mbox{\boldmath $G$}(i\omega_n)$. 
For instance, the self-energy matrix can be described by three functions
 $a_0(i\omega_n)$, $a_3(i\omega_n)$, 
and $a_{\alpha \delta; \lambda \rho}^{\perp}(i \omega_n , i\omega_m)$ :
\begin{align}
 & 
\bm{\Sigma}(i\omega_n) 
 =  a_0(i\omega_n)\, i\omega_n 
 \mbox{\boldmath $1$} 
 + a_3(i\omega_n)\, \xi_d \,\mbox{\boldmath $\tau$}_3 
+ \bm{\Sigma}_{\perp} (i\omega_n) , \\
& \!\!\!
\Bigl\{
 \bm{\Sigma}_{\perp} (i\omega_n) 
\Bigr\}_{\alpha\delta}
\! = 
{1 \over \beta}
\sum_{i\omega_m} 
\sum_{\lambda,\rho}  \,
a_{\alpha \delta; \lambda \rho}^{\perp}(i \omega_n , i\omega_m) 
\nonumber
\\
& \qquad \quad \times
\left\{
 \frac{ \Gamma_{L}^{\phantom{0}} \mbox{\boldmath $\Delta$}_{L}}
 {\sqrt{\omega_n^2 + |\Delta_{L}|^2}} 
+ \frac{ \Gamma_{R}^{\phantom{0}} \mbox{\boldmath $\Delta$}_{R}}
 {\sqrt{\omega_n^2 + |\Delta_{R}|^2}} 
\right\}_{\rho\lambda} .
\label{eq:WT}
\end{align}
The kernel 
$a_{\alpha \delta; \lambda \rho}^{\perp}(i \omega_n , i\omega_m)$
for the off-diagonal part of the self-energy $\bm{\Sigma}_{\perp} (i\omega_n)$
can be shown, using a Ward-Takahashi identity,\cite{ao_ex}
to be expressed in the form,
\begin{align}
&
\!\!
a_{\alpha \delta; \lambda \rho}^{\perp}(i \omega_n , i\omega_m) 
\nonumber\\
 &= 
\sum_{\mu_4} 
\sum_{\mu_2,\mu_3} 
\left\{
\bm{\tau}_3
\right\}_{\alpha\mu_4}
\Gamma_{\mu_4 \delta; \mu_2 \mu_3}
(i \omega_n , i\omega_n\,; i \omega_m , i\omega_m ) 
\nonumber\\
& \qquad \qquad \qquad  \times
\,\Bigr\{
\bm{G}(i\omega_m) 
\,\bm{\tau}_3
\Bigr\}_{\mu_3\rho}
\,\Bigr\{
\bm{G}(i\omega_m) 
\Bigr\}_{\lambda\mu_2} \,.
\end{align}
Here, 
$\Gamma_{\mu_4 \mu_1; \mu_2 \mu_3}
(i \omega_4 , i\omega_1 \,; i \omega_2, i\omega_3 ) $ 
is the vertex function 
in the Nambu representation,
and  $\mu_j$ ($j=1,2,3,4$) represents 
the component of the pseudo spins.
From these properties, it is deduced that, 
if the both couplings and gaps are symmetric 
$\Gamma_{R}^{\phantom{0}}=\Gamma_{L}^{\phantom{0}}$ ($\equiv\Gamma$) and 
$|\Delta_{R}| = |\Delta_{L}|$ ($\equiv\Delta$) 
for the $\pi$-junction $\phi=\pi$, 
the off diagonal part of the
the self-energy vanishes at the impurity site.
It simplifies the Green's function as
\begin{align}
 \!\!\!\!\!\!
\left\{ \mbox{\boldmath $G$}(i \omega_n) \right\}^{-1} 
\Rightarrow &
 \   i \omega_n \!
\left( 1+ 
 \frac{2 \Gamma} {\sqrt{\omega_n^2 + \Delta^2 } } 
- a_0(i\omega_n) \right) \mbox{\boldmath $1$} 
\nonumber \\
&- 
\Bigl( 1+ 
 a_3(i\omega_n) \Bigr)\, \xi_d \, \mbox{\boldmath $\tau$}_3 
.
\label{eq:Green_pi}
\end{align}
Therefore, in this particular case the SC correlation 
cancels out in the impurity site  
$\langle d_{\uparrow}^{\dagger}  d_{\downarrow}^{\dagger} \rangle=0$, 
as can be deduced from eq.\ (\ref{eq:Gdd}). 
Furthermore, if the system 
also has the particle-hole symmetry $\xi_d=0$ additionally,  
the Green's function is simply proportional 
to the unit matrix $\mbox{\boldmath $1$}$.

\section{ limit of $\,|\Delta_R|, \, |\Delta_L| \to \infty$ 
} 
 \label{sec:atomic_limit}

In the limit of  $|\Delta_L| \to \infty$ 
and $|\Delta_R| \to \infty$, the model can be mapped 
onto a simple single-site problem,
\cite{Vecino,AO_YT_Hewson,Affleck}
which can be solved analytically.
In this limit, the Dyson equation eq.\ (\ref{eq:Gdd_inv})
can be rewritten in the form
\begin{align}
& \!\! 
\left\{ \mbox{\boldmath $G$}^{\infty}(i \omega_n) \right\}^{-1} 
=  \  
i \omega_n \mbox{\boldmath $1$} 
-  
\mbox{\boldmath $H$}^0_d
-\mbox{\boldmath $\Sigma$}^{\infty}(i\omega_n) \;,
\label{eq:G_atomic}
\\
&
\mbox{\boldmath $H$}^0_d
=    
\left[ \,
 \begin{matrix}
 \xi_d &  \Delta_d \cr
 \Delta^*_d & -\xi_d
 \end{matrix}
 \, \right]  ,
 \quad \ \  
\Delta_d = 
\Gamma_{L}^{\phantom{0}} e^{i \theta_{L}} 
\!+
\Gamma_{R}^{\phantom{0}} e^{i \theta_{R}} . 
\label{eq:H0_atomic}
\end{align}
The SC correlation penetrates into the dot
from the SC leads via 
 $\Gamma_{L}^{\phantom{0}}$ and  
 $\Gamma_{R}^{\phantom{0}}$. Therefore, 
the amplitude depends on the Josephson phase  
 $\phi = \theta_R - \theta_L$, 
\begin{align}
 \left|\Delta_d \right|^2 
\,= & \ 
\left(\Gamma_R^{\phantom{0}} +\Gamma_L^{\phantom{0}} \right)^2 
- 4 \Gamma_R^{\phantom{0}}\Gamma_L^{\phantom{0}}
\sin^2 \frac{\phi}{2} \;.
\end{align}
The corresponding single-site Hamiltonian is given by
\begin{align}
\mathcal{H}^{\infty}
=& \  \, 
\bm{\psi}_{d}^{\dagger} \,
  \mbox{\boldmath $H$}_d^0 \, 
\bm{\psi}_{d}^{\phantom{\dagger}} 
+
 \frac{2U}{3}\ 
\vec{i}_d \cdot 
\vec{i}_d \;.
\label{eq:H_atom}
\end{align}
The noninteracting part determines 
the energy scale of the Andreev bound state  
$E_A \equiv \sqrt{\xi_d^2 +|\Delta_d|^2}$.
The ground state is a singlet 
for $E_A>U/2$,
while it is a doublet   
for $E_A < U/2$.
The boundary between these two ground states is determined by
the equation $\,U/2=E_A$. 
Since the bound-state energy $E_A$ depends on $\phi = \theta_R - \theta_L$ 
 as      
\begin{equation}
E_A 
=\sqrt{
\xi_d^2 + 
\left(\Gamma_R^{\phantom{0}} +\Gamma_L^{\phantom{0}} \right)^2 
- 4 \Gamma_R^{\phantom{0}}\Gamma_L^{\phantom{0}}
\sin^2 \frac{\phi}{2}
} \;,
\label{eq:E_A}
\end{equation}
the phase transition occurs in between $0<\phi<\pi$, 
if the Coulomb interaction is in 
the region $E_{A\pi} < U/2 < E_{A0}$.
Here, $E_{A0,\pi}$ is the value of $E_A$ at $\phi=0,\pi$.

 \begin{figure}[t]
 \leavevmode 
 \begin{center}
 \includegraphics[width=0.9\linewidth]{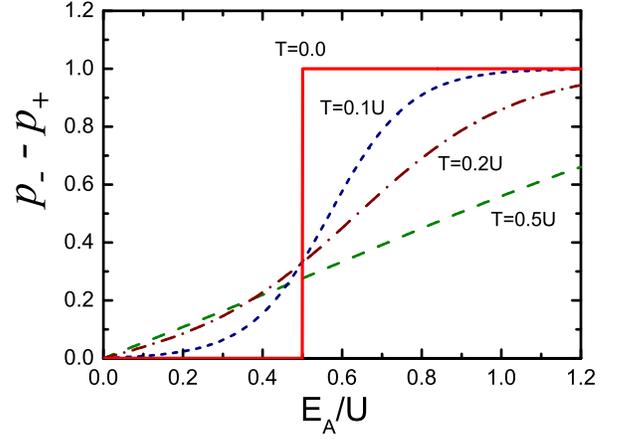}
 \caption{Statistical weight 
 $p_{-} - p_{+}$  as a function of $E_A$ 
for $T/U=0.0$, $0.1$, $0.2$, and $0.5$.
 } 
 \label{fig:weight_atom}
 \end{center}
 \end{figure}

The Green's function can be obtained analytically  
\begin{align}
\mbox{\boldmath $G$}^{\infty}(i\omega_n)
\,= & \sum_{\eta=+,-}
p_{\eta}^{\phantom{0}} \,\, 
\frac{
i \omega_n \mbox{\boldmath $1$} +
\bigl( 1+ \eta \,\frac{U}{2E_A} \bigr)
\mbox{\boldmath $H$}^0_d
}
{(i\omega_n)^2-(\frac{U}{2}+\eta E_A)^2} 
\label{eq:G_inf}
\;,
 \\
  p_{\pm}^{\phantom{0}} \,= & \,\,\frac{1 + e^{-\beta(\frac{U}{2} \pm E_A)}}
 {2 + e^{-\beta(\frac{U}{2}+E_A)} + e^{-\beta(\frac{U}{2}- E_A)}} 
\;.
\end{align}
The equal time correlations, 
$\left.\mbox{\boldmath $G$}(\tau)\right|_{\tau\to -0^+}$, 
is given by  
\begin{align}
 \left[ 
 \begin{matrix}
\langle 
d_{\uparrow}^{\dagger} \,
  d_{\uparrow}^{\phantom{\dagger}} 
\rangle 
& \!
\langle 
d_{\downarrow}^{\phantom{\dagger}} \,
  d_{\uparrow}^{\phantom{\dagger}} 
\rangle 
\cr
\langle 
d_{\uparrow}^{\dagger} \,
  d_{\downarrow}^{\dagger}  
\rangle 
& \!
\langle 
d_{\downarrow}^{\phantom{\dagger}} \,
 d_{\downarrow}^{\dagger} 
\rangle 
 \end{matrix}
 \right]  = \ 
\frac{1}{2}\, \mbox{\boldmath $1$}  +  
\frac{p_{-}^{\phantom{0}} - p_{+}^{\phantom{0}} }{2E_A} \, 
\mbox{\boldmath $H$}^0_d \;.
\end{align}
Note that in the noninteracting case $U=0$, the statistical weight 
$p_{-}^{\phantom{0}} - p_{+}^{\phantom{0}}$
is replaced by $1-2f(E_A)$ with
$f(\epsilon)=\left[e^{\beta\epsilon} +1\right]^{-1}$.  
In Fig.\ \ref{fig:weight_atom}, the weight 
 $p_{-}^{\phantom{0}} - p_{+}^{\phantom{0}}$ is 
plotted as a function of $E_A/U$ for several temperatures.
Specifically at zero temperature, 
$p_{-}^{\phantom{0}} = 1$ and $p_{+}^{\phantom{0}} = 0$, 
for $E_A > U/2$ in the singlet ground state.
It takes the value 
$p_{-}^{\phantom{0}} = p_{+}^{\phantom{0}}=1/2$ 
for $E_A <U/2$ in the doublet ground state. 
Therefore, at $T=0$, the self-energy 
can be expressed in the form 
\begin{align}
\mbox{\boldmath $\Sigma$}^{\infty}(i\omega)
\, = 
\left\{ 
\begin{array}{lll}
\displaystyle
-\frac{U}{2E_A} 
\,\mbox{\boldmath $H$}^0_d &,&  \quad
\displaystyle
E_A > \frac{U}{2}  \\
%
\displaystyle
\frac{U^2}{4}\, 
\frac{i \omega \,\mbox{\boldmath $1$} + 
\mbox{\boldmath $H$}^0_d}{(i\omega)^2-E_A^2} 
\rule{0cm}{0.8cm}
&,& \quad  
\displaystyle
E_A < \frac{U}{2} 
\end{array}
\right.  .
\end{align}

The Josephson current can be calculated  
substituting  eq.\ (\ref{eq:G_inf}) 
into eq.\ (\ref{eq:JR_Joseph}) or (\ref{eq:JL_Joseph}),
\begin{align}
 \langle J \rangle 
 =& \  
 \frac{e}{\hbar} \, 
 \frac{p_{-}^{\phantom{0}} - p_{+}^{\phantom{0}} }{2E_A}  
\ 4\Gamma_R^{\phantom{0}} \Gamma_L^{\phantom{0}}  \sin \phi 
\;. 
\end{align}
At $T=0$, the current can be expressed in the form 
\begin{align}
 \langle J \rangle
\, = 
\left\{ 
\begin{array}{cll}
\displaystyle
\frac{e E_{A0} }{\hbar}\,
\frac{\mathcal{T}\,\sin \phi }{
2 \sqrt{ 1- \mathcal{T} \sin^2 \frac{\phi}{2} } 
} &, &  \quad 
\displaystyle
E_A > \frac{U}{2}  \\
%
\displaystyle
0  \rule{0cm}{0.6cm}
&,  & \quad 
\displaystyle
E_A < \frac{U}{2} 
\end{array}
\right.  .
\label{eq:Josephson_atom}
\end{align}
where $E_{A0}=\sqrt{\xi_d^2+(\Gamma_R + \Gamma_L)^2}$ 
 is the maximum value  of the bound-state energy, and 
 $\mathcal{T} \equiv 
4\Gamma_R^{\phantom{0}} \Gamma_L^{\phantom{0}}/E_{A0}^2$ 
corresponds to a normal transmission probability 
for noninteracting electrons.
The contribution of the Coulomb interaction disappears 
at $T=0$ in the case of the singlet ground state
as discussed by Glazman and Matveev.\cite{GlazmanMatveev}
The Josephson current vanishes for the magnetic doublet ground state 
in the limit of $\,|\Delta_R|, \, |\Delta_L| \to \infty$. 
A finite current flowing in the opposite direction 
in the case of the finite SC gaps,\cite{ShibaSoda}
is caused by the continuous excited states above the gaps.

\end{document}